\documentclass[prl,twocolumn,showpacs]{revtex4}

\usepackage{amsmath}
\usepackage{amsthm}
\usepackage{amssymb}
\usepackage{amscd}
\usepackage[british]{babel}
\usepackage{graphicx}
\usepackage{psfrag}
\usepackage{epsfig}

\def\be{\begin{equation}}
\def\ee{\end{equation}}
\def\bea{\begin{eqnarray}}
\def\eea{\end{eqnarray}}
\def\ba{\begin{array}}
\def\ea{\end{array}}

\begin{document}

\title{The M2/M5 BPS Partition Functions from Supergravity}

\author{Pedro J. Silva$^{1}$\footnote{E-mail address: psilva@ifae.es}}

\affiliation{Institut de Ci\`encies de l'Espai (IEEC-CSIC) and
Institut de F\'{\i}sica d'Altes Energies (IFAE)\\ UAB, E-08193 Bellaterra (Barcelona) Spain}

\date{October 2008}

\thispagestyle{empty}

\begin{abstract}
In the framework of the AdS/CFT duality, we calculate the supersymmetric partition function of the superconformal field theories living in the world volume of either $N$ $M2$-branes or $N$ $M5$-branes. We use the dual supergravity partition function in a saddle point approximation over supersymmetric Black Holes. Since our BHs are written in asymptotically global $AdS_{d+1}$ co-ordinates, the dual SCFTs are in $R\,x\,S^{d}$ for $d=2,5$. The resulting partition function shows phase transitions, constraints on the phase space and allowed us to identify unstable BPS Black holes in the $AdS$ phase. These configurations should correspond to unstable configurations in the dual theory. We also report an intriguing relation between the most general Witten Index, computed in the above theories, and our BPS partition functions.
\end{abstract}

%\pacs{}

\maketitle

%%%%%%%%%%%%%%%%%%%%%%%%%%%%%%%%%%%%%%%%%%%%%%%%%%%%%%%%%%%%%%%%%%%%%%%%%%%%%%%%%%%%%%%%%%%%%%%%
\noindent {\large \bf 1. Introduction}
\vspace{.3truecm}

Partition functions over supersymmetric states $Z_{bps}$, are fascinating objects from which we can extract key information on the corresponding supersymmetric theory, like for example the real number of independent degrees of freedom and the structure of the vacua moduli space. They provide a handle to study extensive properties of the theory including the description of the possible different phases, the order of their transitions, etc.

In general, the calculation of $Z_{bps}$ is not an easy task. If the theory happens to be strongly coupled, things get worse, since not even perturbative approaches can be applied. In the above cases, we may try to calculate other type of objects that hopefully are somehow similar to the unknown partition function. Among these objects we have the so-called Indices of the theory. Their construction is based on representation theory such that, they are by definition invariants of the couplings. Basically, the indices count (modulo some weights) the number of short representations (or BPS rep.), that do not contribute to long representations (general rep.) as the coupling changes. These objects are more easy to calculate, and in certain cases, they are a good approximation to $Z_{bps}$ provided there are no strong cancelations on the characteristic sum within the index.

With the new models of Bager and Lambert \cite{bl}, Gustavsom \cite{gu} and more recently with the ABJM proposal \cite{abjm}, there has been a revival on the research of the $M2/M5$-brane world-volume superconformal field theories (SCFT). Unfortunately, up to date, we are not able to calculate the corresponding supersymmetric $Z_{bps}$, although there are some partial results and vigourous programs currently under development (see \cite{safa} and references there in). On the other hand, there is a proposal for the most general super-conformal Index in four, three, five and six dimensions \cite{sm,s1}. This Indices have been applied to the specific superconformal theories of ${\cal N}=4$ $SU(N)$ SYM in 4D and to the ${\cal N}=6$ $U(N)\times U(N)$ k-level Chern-Simon theory of the ABJM proposal, corresponding to $D3$ and $M2$ brane respectively \cite{sm,s2}.

In the above works, it was found that the index calculated over multiple supergraviton representation of $AdS_{5/4}$ matches exactly the Index calculated on the dual $D3/M2$ SCFT. This result is then interpreted as a non trivial check or support to the AdS/CFT duality in each case. A less fortunate result of these Indices is its relation with the corresponding $Z_{bps}$. It is a fact that these indices show no phase transitions as functions of the different fugacities (or chemical potentials) and therefore do not capture (at least one of the more important features), the associated supersymmetric partition function $Z_{bps}$. Another perspective on this same result comes from the point of view of the dual gravity theory, where these indices are blind to Black Hole (BH) physics.

The main idea of this work, is to obtain the supersymmetric partition function $Z_{bps}$ in the large $N$ limit of the $M2/M5$ world-volume SCFT. This is achieved, studying M-theory supergravity configurations and the relevant $AdS/CFT$ duality. To be more precise, we calculate the supergravity partition function using a saddle point approximation on supersymmetric BH solutions such that $Z_{bps}=e^{-I_{bps}}$. Then, based on the above duality, this object reproduce the corresponding superconformal partition function of the dual theory in the large $N$ limit. We work with supersymmetric M-theory BHs that are asymptotically $AdS_nxS^{11-n}$, $n=4,7$, leaving other types of asymptotic behavior for future research.

To define the supergravity partition function on BPS BH solutions, we need to calculate the supersymmetric Euclidean action $I_{bps}$, in any of the following  ensembles; Micro canonical, Canonical or Grand canonical. We define $I_{bps}$ as the supersymmetric limit of the Euclidean action calculated on non-extremal BHs, in the Grand canonical ensemble. This approach was defined in \cite{yo1,yo2}, and not only provides a natural connection between SCFT BPS statistical mechanics and BPS Euclidean methods in supergravity, but also makes connection with the attractor mechanism and the entropy function of Sen \cite{yo3}.

%%%%%%%%%%%%%%%%%%%%%%%%%%%%%%%%%%%%%%%%%%%%%%%%%%%%%%%%%%%%%%%%%%%%
\vspace{.3truecm}
\noindent {\bf 1.2.~BPS Euclidean actions}\vspace{.3truecm}

In field theory, to define the supersymmetric limit of given partition function, at some point, we have to use that all supersymmetric states saturate a BPS inequality. This equality translates into constraints between the different labeling charges of the associated Hilbert space. To illustrate a general procedure to calculate $Z_{bps}$, let us consider a simple example a Hilbert space characterized by only two labels, say energy $E$ and charge $Q$. We take the BPS bound $E=Q$ \footnote{This type of BPS bound appears in two dimensional supersymmetric models like, {\it e.g.}, the effective theory of $1/2$ BPS chiral primaries of ${\cal N}=4$ SYM in $R\otimes S^3$ (see \cite{Corley:2001zk,Berenstein:2004kk,Caldarelli:2004ig,llm}).}. The Grand canonical partition function $Z$ is a function of two potentials $(\beta,\Omega)$ conjugated to $(E,Q)$ respectively. Define then, the left and right variables $E^\pm= \hbox{$1\over2$}(E\pm Q)$, $\beta_\pm=\beta(1 \pm \Omega)$ such that,
%%%%
\bea Z_{(\beta,\Omega)}=\sum e^{-\beta E + \beta \Omega Q}\,=\sum e^{-\beta_-E_+ - \beta_+E_-}\,.
\eea
%%%%%

The supersymmetric partition function is obtained taking the limit $\beta_+\rightarrow \infty$ while $\beta_-\rightarrow \xi$ (constant). The above limiting procedure takes $T=1/\beta$ to zero, {\it while}
$\Omega$ goes to $\Omega= 1-\xi T + O(T^2)$. Where the new supersymmetric conjugated variable $\xi$ corresponds to the next to leading order in $T$. Note that among all available states, only those that satisfy the BPS bound are not suppress in the sum, giving our resulting supersymmetric partition function
%%%%
\bea
Z_{bps}=\sum_{bps} e^{-\xi Q}=\sum_{Q}d_Qe^{-\xi Q}=\sum_{E}e^{-\xi E}e^{S(E)},\, \eea
%%%%
where the first sum is over all supersymmetric states ($bps$) with $E=J$, in the second sum we have isolated the multiplicity at each $Q$ as $d_Q$ and in the third we solved for $E$ with $S$ equal to the usual entropy.

This limiting procedure can be implemented on supergravity BH solutions, by considering a careful near-to-BPS expansion of the usual Euclidean Action, potentials and charges (the detail explanation and examples can be found in \cite{yo1,yo2}). From the above limiting procedure we are able to define the Euclidean action for BPS BH as a function of the different fugacities $\omega_i$ conjugated to the conserved charges $p^i$. Therefore we can write
\bea
Z_{bps}=e^{-I_{bps}}\,,I_{bps}= \sum \omega_i p^i-S(p^i)\,,i=1,2,...
\label{qsr}\eea
where $Z_{bps}$ stands for the saddle point approximation of the supergravity partition function. There is Another method to calculate the same quantities, using the entropy function of Sen calculated in the near-horizon  of the BPS BH. In this case, the different electric charges correspond to the fugacities, while the function $f$ (Legendre transformed of the entropy function on-shell) is the Euclidean action (see \cite{yo3} for details).

With the above techniques we have been able to compute the partition function over supersymmetric states corresponding to the known BPS BH in $AdS_{4/7}$. Our $Z_{bps}$, shows phase transitions as function of the different fugacities, contrary to the behavior of known Indices. We found small/big BH associated to these phase transitions, where small BH are unstable while big are stable. We also report on a peculiar relation between the partition function $Z_{bps}$ and the associated Index. In short, both objects come equipped with the {\it same constraint} among its fugacities (that also can be read as a constraint among its charges). Presently, we do not understand this issue, but we believe it should play an important role with deep implications to the understanding of superconformal partition function of the world-volume theory of the $M2/M5$.

%%%%%%%%%%%%%%%%%%%%%%%%%%%%%%%%%%%%%%%%%%%%%%%%%%%%%%%%%%%%%%%%%%%%%%%%%%%%%%%%%%%%%%%%%%%%%%%%%%%%%%%
\vspace{.5truecm}
\noindent {\large \bf 2.~M-theory Black holes}
\vspace{.3truecm}

The low energy effective theory of M-theory is conjecture to be ${\cal N}=1$ 11D supergravity. The relevant set up to consider the AdS/CFT duality is to fix boundary conditions such that the asymptotic behavior on each solution is $AdS_{d}xS^{11-d}$ with $d=3,7$, corresponding to UV regime in the near-horizon limit of $N$ $M2$-branes and $M5$-branes respectively. In this framework, M-theory can be  consistently truncated to simple theories, corresponding to the compactification on $S^{11-d}$, defining $d$-dimensional gauge supergravity with R-symmetry group $SO(12-d)$. This theory can be further truncated to the maximal abelian subalgebra of the R-symmetry, corresponding to $U(1)^{m}$ gauge supergravity, where $m=[13-d/2]$. Finally, we can always identify all the $U(1)$ R-charges to get the so called minimal models. Therefore,  M-theory BH solutions can appear at  all the above different levels of truncations. Presently, BPS BHs are known only at the last two levels and it is not clear how generic these solutions are (see \cite{cvetic1} for a study on almost all known solutions).

To calculate the supergravity Euclidean action $I_{bps}$ on BPS Bh, as a function of the different fugacities $(\omega,\phi,..)$ in the Grand canonical ensemble, we follow the procedure introduce in our previous discussions of supersymmetric limits in statistical mechanics. We take the relevant thermodynamic quantities of our family of non-extremal BHs, and study the leading and next-to-leading behavior in a near BPS expansion i.e. as $\beta \rightarrow \infty$. It is a non-trivial fact that we get the following relations,
%%%%
\bea
&&E \rightarrow E_{bps}+ O(\beta^{-2})\,,\quad J\rightarrow J_{bps}+ O(\beta^{-2})\,,\nonumber \\
&&Q\rightarrow Q_{bps}+ O(\beta^{-2})\,,\quad S\rightarrow S_{bps}+ O(\beta^{-1})
\eea while
\bea
\Omega \rightarrow\Omega_{bps}-{w\over\beta }+ O(\beta^{-2})\,,\nonumber\\
\Phi \rightarrow\Phi_{bps}-{\phi\over\beta }+ O(\beta^{-2})\,. \label{limit}
\eea
%%%%
Where the "$bps$" subscript defines the corresponding supersymmetric values of generic angular momentum $J$ and electric charge $Q$, while the next-to-leading terms in the conjugated chemical potentials, define the supersymmetric fugacities.

%%%%%%%%%%%%%%%%%%%%%%%%%%%%%%%%%%%%%%%%%%%%%%%%%%%%%%%%%%%%%%%%%%%%%%%%%%%%%%%%%%%%
\vspace{.3truecm}
\noindent {\bf 2.1~$M2$ case: BPS Black holes in $AdS_4xS^7$}
\vspace{.3truecm}

BPS BHs in asymptotic $AdS_4xS^7$ space-time, are electrically charged, rotating extremal solutions of ${\cal N}=1$ 11D supergravity. They are conjecture to be dual to BPS ensembles of the three dimensional ${\cal N} =8$ SCFT on the world-volume $N$ $M2$-branes, at large $N$. BPS BHs are label by the maximal compact subgroups of the asymptotic isometry group, and therefore are label by it energy $E$, $4d$-angular momentum $J$ and four $U(1)$ R-charges $Q^i$ $i=1,...,4$. Notice that this is precisely the set of labels used to characterize the dual states in the SCFT.

Non-extremal BH solutions of ${\cal N}=8$ $SO(8)$ gauge supergravity truncated to its maximal abelian subalgebra, where found in \cite{cvetic2,caldarelli}, while its supersymmetric limits and thermodynamics are review in \cite{cvetic1}. On these BHs there is an extra parameter labeling the magnetic charge, that is zero in the BPS case. The BPS case only conserves two real supercharges. For simplicity, we only work explicitly the minimal case where all R-charges set equal. Nevertheless, our result are easily generalized to the $U(1)^4$ case.

In the minimal case, the non-extremal electrically charge rotating BH solution comes as a function of three parameter $(m,a,q)$ \footnote{here we follow the conventions of \cite{caldarelli} with $AdS_4$ radius set equal to 1.}. The thermodynamic potentials are,
\bea
&&\beta = {4\pi(r_+^2-a^2)\over{r_+[1+a^2+3r+^2-(a^2+q^2)/r_+^2]}}\,, \nonumber\\
&&\Omega = {a(1+r_+^2)\over (r_+^2+a^2)},\quad
\Phi={qr_+\over(r_+^2+a^2)}\,,
\eea
where $r_+$ is a function of $(m,a,q)$ corresponding to the radial position of the outer horizon. The three different charges and entropy $S$ are
\bea
&&E = {m\over(1-a^2)^2}\,,\quad J = {am\over (1-a^2)^2}\,,\nonumber\\
&&Q={q\over4(1^2-a^2)}\,,\quad S={\pi(r_+^2+a^2)\over 1-a^2}
\eea
while the corresponding Euclidean can be written as
\bea I=\beta\,E-\beta\,\Omega\,J + 4\beta\,\Phi\, Q- S(E,J,Q)\,. \label{qsr1}\eea

The extremal BH in the BPS regime is obtained imposing the BPS constrain
\be
E=J+4Q\,,\label{bps}
\ee
together with the requirement that no closed time-like (CTC) curves are found outside the horizon (see \cite{cvetic1} for details). This last two conditions, reduce the total number of independent degree of freedom to only one.

If we oxidate the BPS BH to 11D, using co-ordinates that are asymptotically static to $AdS_4xS^7$, the solution rotates in both factors, $AdS_4$ and $S^7$, in all possible directions, with velocities equal to the velocity of light, i.e. $\Omega_{bps}=1, \Phi_{bps}=1$.

To obtain the correct expressions for the BPS partition function and to define the different chemical potentials, we have to calculate the leading and next to leading term in a near BPS expansion on the above family of non-extremal BHs. To do this, we chose a off-BPS parameter $\mu$ such that,
\bea
m^2=a(1+a)^4+\mu\,,\quad q=\sqrt{a}(1+a) \label{mu-bps}
\eea
where $\mu=0$ reproduces the BPS BH. The corresponding expansion of the charges and entropy in terms of $\mu$ gives
\bea
E={\sqrt{a}\over(1-a)^2}+O(\mu)\,, J={\sqrt{a}a\over(1-a)^2}+O(\mu)\,, \nonumber \\
Q={\sqrt{a}\over 4(1-a)}+O(\mu)\,, S={\pi a\over (1-a)}+O(\sqrt{\mu})\,,
\eea
while the expansion of its conjugated potentials gives;
\bea
&&\beta={\sqrt{2}\pi a^{3/4}(1-a)\over\sqrt{a^2 +6a+1}}{1\over\sqrt{\mu}} +O(0)\,,\nonumber\\
&&\Omega=1-\frac{2\sqrt{2}(1-a)}{a^{1/3}\sqrt{a^2 +6a+1}}\sqrt{\mu} +O(\mu)\,, \nonumber\\
&&\Phi=1-\frac{\sqrt{2}(1-a)}{a^{1/3}\sqrt{a^2 +6a+1}}\sqrt{\mu} +O(\mu)\,.
\eea
From the above expansion, using eqn. (\ref{limit}), we can read off the parametric form of the BPS charges, entropy and fugacities,
\bea
&&J_{bps}={\sqrt{a}a\over(1-a)^2}\,, \quad\omega={4\pi\sqrt{a}(1-a)^2\over(1+a)\sqrt{a^2 +6a+1}}\,,\nonumber  \\
&&Q_{bps}={\sqrt{a}\over 4(1-a)}\,, \quad\phi={2\pi\sqrt{a}(1-a)^2\over(1+a)\sqrt{a^2 +6a+1}}\,,\nonumber\\
&&E_{bps}={\sqrt{a}\over(1-a)^2}\,, \quad S_{bps}={\pi a\over (1-a)} \,,
\label{cp4}
\eea
that allows us to write the Euclidean action as a function of $(\omega,\phi)$ as follows,
\be
I_{bps}=\omega J_{bps}+4\phi \, Q_{bps}-S_{bps}\,.
\ee

Finally we can use the BPS equation (\ref{bps}), to rewrite $I_{bps}$ in terms of the fugacities related to $(E,Q)$ as follows,
\bea
I_{bps}=\xi\, E_{bps}-4\mu\, Q_{bps}-S_{bps},\,,
\label{I4b}
\eea
with $\xi=\omega\,,\;\mu=\omega-\phi$.

%%%%%%%%%%%%%%%%%%%%%%%%%%%%%%%%%%%%%%%%%%%%%%%%%%%%%%%%%%%%%%%%%%%%%%%%%%%%%%%%%%%%%%%
\vspace{.3truecm}
\noindent {\it 2.1.1~$M2$ Phase transitions, stable/unstable Bh and constraints}
\vspace{.3truecm}

The Euclidean action for the $M2$-BH written in terms of the parameter $a$ is given by the expression
\be
I_{bps}={\pi a(a^2+8a-1)\over(a-1)(a^2+6a+1)}\,,
\ee
where $a$ runs on the interval $(0,1)$ and the BH radius is $r=\sqrt{a}$. The above is our working expression summarizing our results for the Euclidean action in the Grand canonical ensemble of BPS BHs.

\begin{figure}
\includegraphics{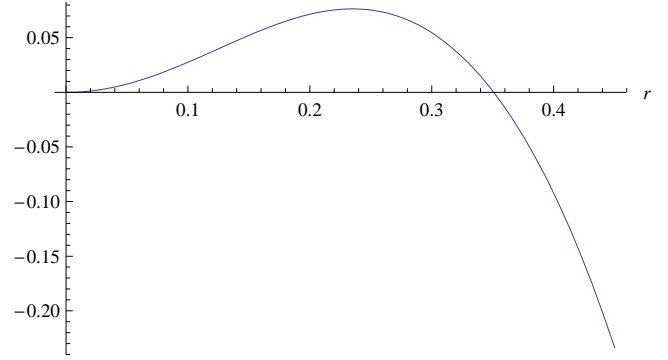}
\caption{Plot of the Euclidean action of the BPS BH
as a function of the BH radius.\label{I4}}
\end{figure}

The corresponding partition function shows a clear phase transition between two phases (see fig \ref{I4}). These non-coexisting phases should correspond to a sea of supergravitons in $AdS_4$ and a BH in $AdS_4$. In terms of the SCFT degrees of freedom, both phases are very different, since the action scales like $N^{3/2}$ in the BH phase in contrast to the scale $N^{0}$ characteristic of the $AdS$ phase. The phase transition is of first order as can be seen from the calculation of the different susceptibilities.

\begin{figure}
\includegraphics{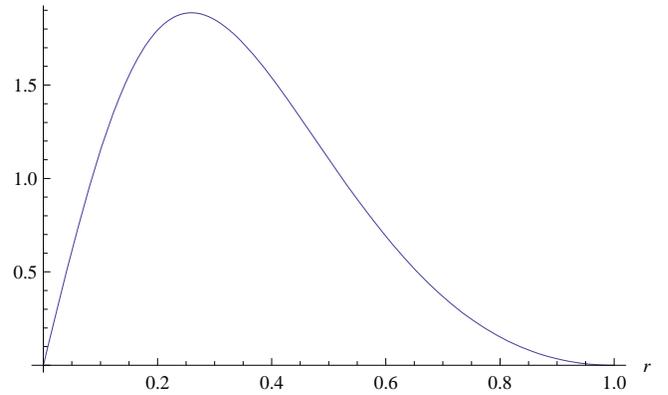}
\caption{Plot of the fugacity $\xi$ as a function of the BH radius. It can be seen that given one value of $\xi$ there correspond two solutions for the BH radius $r$, a small and big BH respectively.\label{F4}}
\end{figure}

The role of the unstable BPS BHs in the $AdS$ phase is similar to the that of small BHs at finite temperature in $AdS$. They are unstable saddle points of the Euclidean action, and should therefore be dual to unstable configurations in the SCFT (see \cite{ho} for a similar phenomenology on the $D3$-brane case.). To better illustrate this point, let us plot in fig. (\ref{F4}) the fugacity $\xi$ (potential conjugated to the Energy) as a function of $r$. We can indeed see, that $\xi$ is double value showing two branches. The first/second branch corresponds to small/big BHs. Small BHs are found between the origin and the maximum value of $\xi$, while big BHs are found from this maximum until the end.

As a last comment, it is important to stress that these BPS BHs are constraint systems. There is only one free degree of freedom (that we parameterized with $a$), while in principle, from naive expatiations on the dual SCFT there should be two. Recalled that our BH solutions are labeled by three charges $(E,J,Q)$, that we have the BPS constraint of eqn. (\ref{bps}) and a causality constraint to eliminate CTC. This last extra constraint adopts a complicated form in terms of the BH charges (we give it implicitly in eqn.(\ref{mu-bps})), but is particularly simple when written in terms of the conjugated fugacities, giving
\bea
\xi=2\mu\quad\hbox{equivalent to}\quad\omega=2\phi\,.
\label{cons1}\eea
We will come back to this issue when discussing the SCFT partition function and its Index in section $3$.

%%%%%%%%%%%%%%%%%%%%%%%%%%%%%%%%%%%%%%%%%%%%%%%%%%%%%%%%%%%%%%%%%%%%%%%%%%%%%%%%%%%%%%%%%%%%%%%%%%%
\vspace{.3truecm}
\noindent {\bf 2.2.~$M5$ case: BPS Black holes in $AdS_7xS^4$}
\vspace{.3truecm}

BPS BHs in asymptotic $AdS_7xS^4$ space-time, are electrically charged, rotating extremal solutions of ${\cal N}=1$ 11D supergravity. They are conjecture to be dual to BPS ensembles of the six dimensional ${\cal N} =(2,0)$ SCFT on the world-volume $N$ $M5$-branes, at large $N$. BPS BHs are label by the maximal compact subgroups of the asymptotic isometry group, and therefore are label by it energy $E$, three $7d$-angular momenta $J_I$ $I=1,2,3$ and two $U(1)$ R-charges $Q^i$ $i=1,2$. Again, this is precisely the set of labels used to characterize the dual states in the SCFT.

Non-extremal BH solutions of ${\cal N}=4$ $SO(5)$ gauge supergravity truncated to its maximal abelian subalgebra, where found in \cite{cvetic3,chow} and its supersymmetric limits and thermodynamics are review in \cite{cvetic1,chow}. The general non-extremal BH solution depending on all six charges in not known, where the known solution have either same angular momenta and different electric charges \cite{cvetic3}, or same electric charges and unequal angular momenta \cite{chow}. For simplicity, we only work out the minimal case, where all R-charges are set equal to $Q$ and all the angular momenta are set equal $J$. Nevertheless, our result are easily generalized to the other available cases. All known BPS cases only conserve two real supercharges.

In this minimal case, the non-extremal electrically charge rotating BH solution comes as a function of three parameter $(m,a,d)$ \footnote{here we follow the conventions of \cite{chow} with $AdS_7$ radius set equal to 1.}. The thermodynamic potentials are,
\bea
&\beta = {2\pi r_+[(r_+^2+a^2)^3+q(r_+^2-a^3)]\over -q^2 + 3r_+^2(1 + r_+^2)(a^2 + r_+^2)^2-(a^2 + r_+^2)^3+2q(a^3 + r_+^4)}\,,& \nonumber\\
&\Omega = {a[(r_+^2+a^2)^2(1+r_+^2)+q(r_+^2-a)] \over (r_+^2+a^2)^3+q(r_+^2-a^3)}\,,&\nonumber \\
&\Phi ={\pi^2\cosh(d)\sinh(d)\over (1-a^2)^3}\,,&
\eea
where $q=2m\sinh(d)^2$ and $r_+$ is a function of $(m,a,d)$ corresponding to the position of the outer horizon. The three different charges and entropy  $S$ are
\bea
&E= {m \pi^2 [-5-a^2+(-8 + 11a^2+12a^3+3a^4)\sinh(d)^2]\over 8 (1 - a^2)^4}\,,&\nonumber\\
&J= {am\pi^2(-\cosh(d)^2 +a(1+a)^2\sinh(d)^2)\over (1 - a^2)^3}\,,&\nonumber\\
&Q={\pi m\sinh(d)\cosh(d)\over 2(1-a^2)^3}\,,&\nonumber\\
&S={\pi^3[(a^2 + r_+^2)^3 + q(ro^2-a^3))\over 4(1-a^2)^3r_+}\,.&
\eea
Then, in the Grand canonical ensemble the Euclidean action $I$ can be written as
\bea I=\beta\,E-3\beta\,\Omega\,J - 2\beta\,\Phi\, Q- S(E,J,Q)\,. \label{qsr2}\eea

In the above conventions, supersymmetric BHs are obtained imposing the BPS constrain
\be
E+3J-2Q=0\,,\label{bps2}
\ee
together with the requirement that no closed time-like curves (CTC) are found outside the horizon (see \cite{cvetic1} for details). This last two conditions, reduce the total number of independent degree of freedom to only one. As in the $M2$ case, the oxidated BPS BH in 11D (in co-ordinates that are asymptotically static to $AdS_7xS^4$), is rotating in both factors, $AdS_7$ and $S^4$, with velocities equal to the velocity of light, i.e. $|\Omega_{bps}|=1$, $\Phi_{bps}=1$.

To obtain the corresponding BPS partition function, we calculate the leading and next to leading term in a near BPS expansion on the family of non-extremal BHs. The off-BPS parameter is $\mu$ such that,
\bea
m={3qa(3a-2)\over2}+\mu\,, q={8a^3(a-1)^3\over(1-3a)^2} \label{mu-bps2}
\eea
where $\mu=0$ reproduces the BPS BH. The corresponding expansion of the charges and entropy in terms of $\mu$ gives
\bea
E={\pi^2a^3(-4+11a-6a^2+3a^3)\over(1+a)^4(1-3a)^2} +O(\mu)\,,\nonumber\\
J=-{\pi^2a^4(1-6a+a^2)\over (1+a)^4(1-3a)^2} +O(\mu)\,,\nonumber\\
Q={2\pi^2a^3\over(1+a)^3(3a-1)} +O(\mu)\,,\nonumber\\
S={2\pi^3a^4\sqrt{3-a}\over(1+a)^3\sqrt{(1-3a)^3}}+O(\sqrt\mu)\,,
\eea
while the expansion of its conjugated potentials gives;
\bea
&\beta={\pi[(r_+^2+a^2)^3+q(r_+^2-a^3)]\over Br_+[2q+3(a^2+r_+^2)(1+a^2+2r_+^2)]}{1\over\sqrt{\mu}} + O(0)\,,\qquad \quad&\nonumber\\
&\Phi=1-\frac{Ba^2[(r_+^2+a^2)(a^2-2r_+^2)+q a]}{r_+^2[(r_+^2+a^2)^3-q(r_+2-a^3)]}\sqrt{\mu} +O(\mu)\,,\qquad \quad&\nonumber\\ &\Omega=-1+\frac{B(1+a)(a^4+2a^3+4a^2r_+^2+2ar_+^2+3r_+^2+q)}{[(r_+^2+a^2)^3-q(r_+2-a^3)]}\sqrt{\mu}+\nonumber \\
&\hspace{4truecm} +\, O(\mu)\,.&
\eea
Where $r_+^2=a^2(3-a)/(1-3a)$ is the BPS BH radius and $B$ a polynomial in $a$. From the above expansion, using eqn. (\ref{limit}), we can read-off the parametric form of the BPS charges, entropy and fugacities
\bea
&E_{bps}={\pi^2a^3(-4+11a-6a^2+3a^3)\over(1+a)^4(1-3a)^2}, \, S_{bps}={2\pi^3a^4\sqrt{3-a}\over(1+a)^3\sqrt{(1-3a)^3}} \,,&\nonumber  \\
&J_{bps}=-{\pi^2a^4(1-6a+a^2)\over (1+a)^4(1-3a)^2}, \,\omega={-8\pi a^2(1+a)\sqrt{1-3a}\over\sqrt{(3-a)a^2}[3+a(19a-10)]}\,,&\nonumber  \\
&Q_{bps}={2\pi^2a^3\over(1+a)^3(3a-1)}, \,\phi={6\pi a^2(1+a)\sqrt{1-3a}\over\sqrt{(3-a)a^2}[3+a(19a-10)]}\,.&\nonumber\\
\label{cp7}\eea
These expressions allow us to write the Grand canonical Euclidean action,
\be
I_{bps}=3\omega J_{bps}+2\phi \, Q_{bps}-S_{bps}\,,
\ee
as a function of $(\omega,\phi)$ only. As we proceeded in the $M2$ case, we use the BPS equation (\ref{bps2}), to rewrite $I_{bps}$ in terms of the fugacities related to $(E,Q)$ as follows,
\bea
I_{bps}=\xi\, E_{bps}-2\mu\, Q_{bps}-S_{bps}\,,
\label{I7b}\eea
with $\xi=-\omega\,,\;\mu=-(\omega+\phi)$.

%%%%%%%%%%%%%%%%%%%%%%%%%%%%%%%%%%%%%%%%%%%%%%%%%%%%%%%%%%%%%%%%%%%%%%%%%%%%%%%%%%%%%%%%
\vspace{.3truecm}
\noindent {\it 2.2.1.~$M5$ Phase transitions, stable/unstable BH and constraints}
\vspace{.3truecm}

The parametric form of the Euclidean action for the $M5$-BH is
\bea
&I_{bps}={2\pi^3a^5\{-3 + a[3 + (31 - 7a)a]\}\sqrt{1-3a}\over (1-3a)^2(1 + a)^3(3 +a(-10+19a))\sqrt{((3-a) a^2}}\,,&
\eea
where $a$ runs on $(-1,0)$. This is our final expression for the Euclidean action in the Grand canonical ensemble for the BPS BHs.

\begin{figure}
\includegraphics{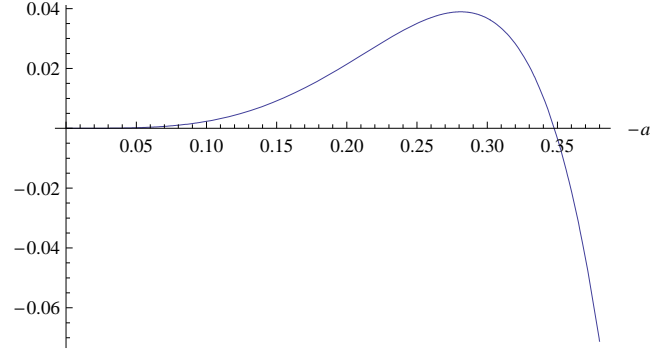}
\caption{Plot of the Euclidean action of the BPS BH
as a function of the parameter $-a$.\label{I7}}
\end{figure}

As in the $M2$ case, the corresponding partition function shows a phase transition between two phases (see fig \ref{I7}). These non-coexisting phases should correspond to a sea of supergravitons in $AdS_7$ and a BH in $AdS_7$. In terms of the SCFT degrees of freedom, the action scales like $N^{3}$ in the BH phase in contrast to the scale $N^{0}$ characteristic of the $AdS$ phase. The phase transition is of first order as can be seen from the calculation of the different susceptibilities.

\begin{figure}
\includegraphics{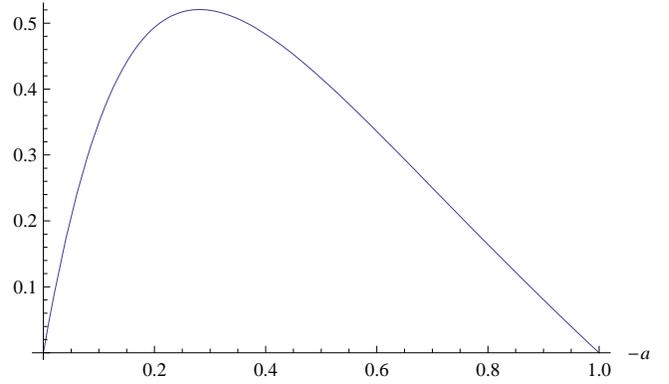}
\caption{Plot of the fugacity $\xi$ as a function of -$a$. Given one value of $\xi$ there correspond two different BHs, with different radius; small and big BHs.\label{F7}}
\end{figure}

We have the same situation as in the $M2$-BH, where the unstable BPS BHs (small BHs) in the $AdS$ phase corresponds to unstable saddle points of the Euclidean action, and should therefore be dual to unstable configurations in the SCFT. In fig. \ref{F7}, the potential conjugated to the Energy $\xi$ is ploted as a function of a. Here it can be seen that it is double valued function of $a$, showing two branches of small/big BHs. Small BHs are found between the origin and the maximum value of $\xi$, while big BHs are found from this maximum until the end.

To closed this section, this system also shows an extra constraint over the BPS relation of (\ref{bps2}). There is only one free degree of freedom (that we parameterized with the letter $a$), while in principle, from naive expatiations on the dual SCFT there should be two. This extra constraint adopts a complicated form in terms of the BH charges (we give it implicitly in eqn. (\ref{mu-bps2})), but is particularly simple when written in terms of the conjugated fugacities, giving
\bea
\xi=4\mu\quad\hbox{equivalent to}\quad \omega={4\over3}\phi\,.
\eea
In the next section, we will elaborate on this and the $M2$-BH constraint of equation (\ref{cons1}) showing its coincidence with the constraint that characterized the most general index of these SCFT.

%%%%%%%%%%%%%%%%%%%%%%%%%%%%%%%%%%%%%%%%%%%%%%%%%%%%%%%%%%%%%%%%%%%%%%%%%%%%%%%%%%%%%%%%%%%%%%%%
\vspace{.5truecm}
\noindent {\large \bf 3.~ Indices and Partition functions}
\vspace{.3truecm}

The field theory calculation of the $M2/M5$ SCFT partition functions in the large $N$ limit are unknown. Our incomplete understanding is due to the fact that we do not have a good description of the world-volume theory of multiple $M2/M5$. Even in the case of the ABJM proposal, where there is a candidate for the SCFT at level $k$ the calculation is presently out of our possibilities (see \cite{safa} for results on this direction.). Nevertheless at least formally, it should be a function of all the potentials conjugated to the charges that label our supersymmetric states in the ensemble.

The situation is better when considering supersymmetric Witten indices. In \cite{sm,s1}, it was possible to define and to give a prescription of how to calculate the most general Witten Index $I^w$ for SCFT in $D=3,4,5,6$. This Index has already been computed in the AdS/CFT framework, showing perfect agrement on both sides of the duality \cite{sm,s2} for the cases where we have a realization the SCFT.

To define the most general $I^w$ in SCFT, we first choose an arbitrary conjugated pair of supersymmetric generators $(Q,S)$, that define the unbroken supersymmetries of our BPS ensembles of states. Then, the index is written as a weighted sum over the Hilbert subspace of states that are annihilated by $(Q,S)$. Since all our BPS states transform in a irreducible representation of the subalgebra of the superconformal algebra that commutes with our chosen pair $(Q,S)$, it is clear that this maximal commuting subalgebra (MCS) plays a key role in the definition of $I^w$. The corresponding trace formula for $I^w$ is
\bea
I^w=Tr_H[(-1)^F\exp(-\beta\{Q,S\}+G)]\,,
\eea
where we traced over the full Hilbert space $H$, $F$ is the Fermion number operator, $G$ is an element of MCS.

The resulting index is independent of $\beta$ and is therefore label by elements of the MCS only (see \cite{s1} for a detail explanation on all the above). Due to this last point, $I^w$ is in general a function of less parameters than the one needed to label a BPS state and therefore is defined as a constrained function on the general phase space of the theory.

In the following, we will extract these constraints in terms of the natural fugacities conjugated to the labeling charges of our BPS states. We work this out for the particular cases of the 3D SCFT with R-symmetry $SO(8)$ and for the 6D SCFT with R-symmetry $Sp(4)$. The resulting constraints should corresponds to the $M2/M5$-SCFT constraints that takes place in the definition of $I^w$.

%%%%%%%%%%%%%%%%%%%%%%%%%%%%%%

For the large $N$ limit case of $N$ $M2$-branes, $I^w$ can be calculated over multi-gravitons states in $AdS_4xS^7$. In this situation, we first calculate the index over each graviton representation $I_{sp}$, to then sum over all single gravitons and multi-gravitons \footnote{We basically follow the notation of \cite{s1}.}. The index on each graviton rep. goes as
\be
I_{Rn}=Tr_{bps}\left[(-1)^F \exp(-\rho E+\sum_{i=1}^3\gamma_iH^i)\right]\,,
\ee
where $(E,H^i)$ are the Cartan charges of the bosonic subgroup $SO(2,1)xSO(6)$ of the MCS. Its relation to the full set of superconformal Cartan charges $(e,j,h^i)$ is
\bea
E=e+j\,,\quad H^i=h^{i+1}
\eea
where $i$ runs over $2,3,4$ and $(e,j,h_i)$ stands for energy, angular momentum and R-charge respectively.
The BPS constraint in these conventions turns out to be
\be
e-j-h^1=0\,.
\ee
To find the form of the constraint among the different fugacities, we write
\be
\eta j +\sum_{i=1}^4 \lambda_i h^i = -\rho E+\sum_{i=1}^3\gamma_iH^i
\ee
from which we get, after using the above BPS relation; $\rho=-\lambda_1$, $\gamma_{i}=\lambda_{i+1}$ for $i=1,2,3$ and $\eta-\lambda_1=-\rho$. That allow us to find the corresponding constraint appearing in $I^w$,
\be
2\eta-\gamma_1=0\,.
\ee
Rewriting the above constrain in terms of the $U(1)^4$ gauge supergravity in 4D of previous sections, gives
\be
2\omega-{1\over4}\sum_{i=1}^4\phi_i=0\,,
\label{M2const}\ee
that in the case of equally R-charged ensemble (i.e. $\phi_i=\phi$ for all $i$), gives
\be
2\omega=\phi\;\longleftrightarrow\;\xi=2\mu\,.
\ee
Where in the last expression, we have used the fugacities of eqn. (\ref{I4b}). Therefore, we have found that:

\vspace{.3truecm}
\hspace{.5truecm}\emph{The constraint appearing in the Witten index of our $M2$-SCFT at large $N$ is exactly
the same constraint that appears in known $M2$-BHs}.
\vspace{.3truecm}

%%%%%%%%%%%%%%%%%%%%%%%

For the $M5$ case, we proceed in a similar fashion. First we write the corresponding trace formula for $I_{Rn}$ in $AdS_7xS^4$,
\be
I_{Rn}=Tr_{bps}\left[(-1)^Fe^{(-\rho E+\sum_{i=1,2}\gamma_iH^i+\zeta K^1)}\right]\,,
\ee
where $(E,H^1,H^2,K^1)$ are the Cartan charges of the bosonic subgroup $SO(5,1)xSp(2)$ of the MCS. Its relation to the full set of superconformal Cartan charges $(e,h^I,k^0,k^1)$ with $I=1,2,3$ is
\bea
&&E=3e+h^1+h^2-h^3\,,\quad H^1=h^1-h^2\,,\nonumber\\
&&H^1=h^2+h^3\,,\quad K^1=k^1\,,
\eea
where $(e,h^I,k^0,k^1)$ are respectively energy, angular momenta and R-charges.
The BPS constraint in these conventions turns out to be
\be
e-h^1-h^2+h^3-4k^0=0\,.
\ee
To find the form of the constraint among the different fugacities, we write
\be
\sum_{I=1}^3\eta_I h^I +\sum_{i=0,1}\lambda_i k^i = -\rho E+\sum_{l=1,2}\gamma_lH^l+\zeta K^1\,.
\ee
From the above equation and the BPS relation, we get that the constraint in this case is
\be
\eta_1+\eta_2-\eta_3-\gamma_0=0\,.
\ee
Rewriting the above constrain in terms of the $U(1)^2$ gauge supergravity in 7D of previous sections, gives
\be
\sum_{I=1}^3\omega_I+{1\over2}\sum_{i=1}^2\phi_i=0\,.
\label{M5const}\ee
In the case of equal angular momenta and equal R-charge (i.e. $\omega_I=\omega$ and $\phi_i=\phi$ for all $I,i$), the constraint reduces to
\be
3\omega-4\phi=0\;\longleftrightarrow\;\xi=4\mu\,,
\ee
where in the last expression we have used the fugacities of eqn. (\ref{I7b}). Therefore we have found that:

\vspace{.3truecm}
\hspace{1truecm}\emph{The constraint appearing in the definition of the Witten index of the $M5$-SCFT at large $N$, is the same constraints that appears in known $M5$-BHs.}
\vspace{.5truecm}

%%%%%%%%%%%%%%%%%%%%%%%%%%%%%%%%%%%%%%%%%%%%%%%%%%%%%%%%%%%%%%%%%%%%%%%%%%%%%%%%%%%%%%%%%%%%%%%%%%%%%%%%%%
%\vspace{.5truecm}
\noindent {\large \bf 5.~Discussion}
\vspace{.3truecm}

In this work, we have used the AdS/CFT duality to define the supersymmetric partition function of the $M2/M5$ SCFT at large $N$, $Z_{bps}$. We have calculated the M-theory supergravity partition function as a saddle point approximation on supersymmetric BH. In turns, to calculate this supergravity partition function, we need the Euclidean action on these BPS BHs, as a function of all the fugacities conjugated to all the different charges that label our BPS states.

The Euclidean action on the BPS BHs is defined as the supersymmetric limit of the Euclidean action of non-extremal BHs in M-theory \cite{yo1,yo2}. There is another method, base on the attractor mechanism and Sen's entropy function, that is equivalent as shown in \cite{yo3}.

Our result are summarized by eqns. (\ref{cp4},\ref{I4b}) for the $M2$-case, and eqns. (\ref{cp7},\ref{I7b}) for the $M5$-case. The resulting partition functions show clear phase transitions as a function of the fugacities. In the $AdS$ phase we found the existence of small BHs that are unstable, representing local maximums of the Euclidean action and therefore should be dual to unstable configurations in the SCFT (see \cite{ho} for similar but different behavior in the finite temperature for the case $D3$-branes.).

We also found that all known BPS BHs in M-theory are constraint, and that these constraints are found in the definition of the most general Witten Index $I^w$ of the corresponding SCFT. These constraints are given in eqn. (\ref{M2const},\ref{M5const}) for the $M2$ and $M5$ cases respectively. The above observation is an intriguing fact, since our BHs should be related to the dual partition function and a priory; there is no reason for this relation with the Index. We believe that there is a deep reason for the above relation, that should have consequences in the desiderate calculation of the BPS partition function (using field theory approaches), and also on the issue of finding the most general BPS BH in M-theory.

The above result is also verified for the $D3$-brane case, where the constraint of BPS BH in $AdS_5$ is the same constraint of the Index on the dual ${\cal N}=4$ $4D$ SYM theory \cite{yo2}. One interesting application was studied in \cite{yo4}, where CFT information regarding the extra-constraint, was used to treat the problem of "BH uniqueness in 5D gauge supergravity" .

%%%%%%%%%%%%%%%%%%%%%%%%%%%%%%%%%%

\textsl{We thank M. Panareda for her infinite patience and support. This work was partially funded by the Ministerio de Educacion y Ciencia under grant CICYT-FEDER-FPA2005-02211 and CSIC under the I3P program.}

%%%%%%%%%%%%%%%%%%%%%%%%%%%%%%%%

%%%%%%%%%%%%%%%%%%%%%%%%%%%%%%%%
\end{document}